\documentclass[twocolumn]{IEEEtran}
\IEEEoverridecommandlockouts
\usepackage[pdftex]{graphicx}
\usepackage{epstopdf}
\usepackage{amsmath}
\usepackage{stfloats}

\title{
{
~\vspace{-40pt}
\normalsize PREPRINT; to appear open access in IEEE Transactions on Power Systems; accepted January 2016} \\
\vspace{40pt}
Obtaining Statistics of Cascading Line  Outages Spreading in an Electric  Transmission Network from Standard Utility Data
 }

\author{ Ian~Dobson, %~\IEEEmembership{Fellow~IEEE}, 
Benjamin A. Carreras, David E. Newman, Jos\'e M. Reynolds-Barredo%,
\thanks{\looseness=-1
I. Dobson is with ECpE dept., Iowa State  University, Ames IA 50011 USA;
 dobson@iastate.edu.
 B. A. Carreras is with BACV Solutions Inc., Oak Ridge, TN USA; email bacarreras@gmail.com, D. E. Newman  is with Physics Dept.,
University of Alaska, Fairbanks AK 99775 USA; email ffden@uaf.edu, J. M. Reynolds-Barredo is with Universidad Carlos III de Madrid, Spain; email jmrb2002@gmail.com.
Funding in part  by NSF grant 
CPS-1135825 is gratefully acknowledged.
This work is licensed to IEEE under the Creative Commons Attribution 3.0 United States License (CC BY 3.0 US).
}}

\begin{document}

%\markboth{to be submitted to .}{Shell}

\maketitle

\begin{abstract}
We show how to use standard  transmission line outage historical data to obtain the network topology 
in such a way  that cascades of line outages can be easily located on the network.
Then we obtain  statistics quantifying how cascading outages typically spread on the network.
Processing real outage data  is fundamental for understanding cascading 
and for evaluating the validity of  the many different models and simulations that have been proposed
for cascading in power networks.
\end{abstract}


\section{Introduction }

Complicated series of cascading outages in the transmission network occasionally cause blackouts. 
These large cascading blackouts are rare, but of substantial risk since their impact is high \cite{NewmanREL11,HinesEP09}.
In cascading, initial outages  propagate and progressively spread across the network, and, if there are many outages, load is shed and there is a blackout.
The initial outages can be random failures due to  many different causes, including weather, animals, equipment malfunction, earthquakes, operational errors and malicious attacks. The subsequent spreading of the outages beyond the initial outages in a cascade of dependent outages is  complicated and includes many ways in which 
multiple previous outages or a common cause such as weather can weaken the transmission network to make further outages more likely.

\begin{figure}[t]
  \centering
 \includegraphics[width=0.5\columnwidth]{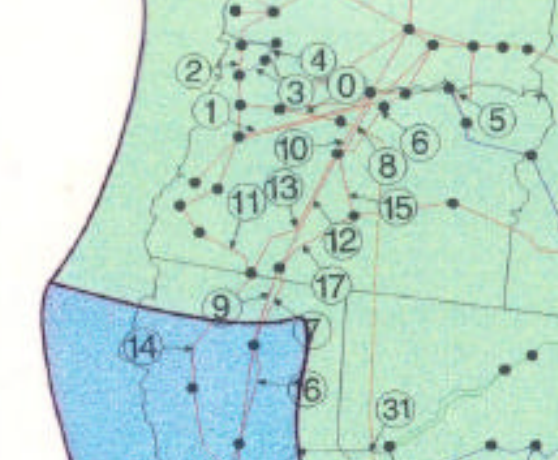} 
  \caption{\looseness=-1 Initial portion of  Western interconnection blackout of August~10~1996. Numbers show the order of the outages. Figure is  extracted from \cite{NERC02}.}
  \label{blackout1996}
\end{figure}

\looseness=-1
 To  motivate our study, we first consider how the transmission line outages spread in  the August 10 1996 Western interconnection blackout.
The NERC blackout report \cite{NERC02} shows the initial spread of the cascading as reproduced in Fig.~\ref{blackout1996}.
The numbers show the order of the outages.
It is clear that the outages propagate to other outages both near and far in the network and that the 
total extent of the cascade spreading can be large.
In particular, the first 18  line outages of the blackout occur on the network formed in section \ref{forming}
that is a subnetwork of the Western interconnection,
so we located the 18  outages on this network.
One way to measure the distance between two line outages counts the minimum number of buses in 
a path in the network joining the two lines.
For example, two lines with a common bus are a distance one apart.
In the case of the  first 18  line outages of the blackout, we find that 
the distance 
between successive line outages ranges from 2 to 6 buses and has a mean value of 3.2.
The maximum extent of the first 18 cascading  outages  is 4 buses away from the initial outage.

\looseness=-1
This example of the August~10~1996 blackout shows the initial 
spread of one blackout and shows what can possibly happen.
 However, 
one cannot draw general conclusions about how cascades typically spread from one sample.
Indeed, the August 10  1996 blackout is one of the more serious blackouts that has ever occurred in the Western interconnection, whereas
the most common cascades, by careful design and operation of the power system, are that an initial outage occurs and either no outages
or only a few outages follow.
In order to account for successful mitigation as well as failures of mitigation, the assessment and mitigation of cascading risk must account for  cascades of all sizes.

\looseness=-1
The detail of the spreading of  cascading outages can be studied either by analyzing the complexities of particular blackouts after they occur
\cite{ManiIREP04, KosterevPS96}, or by simulating some subset of the mechanisms for cascading 
\cite{ChenEPES05,HardimanPMAPS04,RiosPS02,CarrerasCH04}. These approaches are very useful both in understanding cascading blackouts and 
suggesting ways to mitigate particular mechanisms of cascading failure.
The spatial correlation of Euclidean distance between outages is computed in \cite{DaqingSR14} for the July 2 and August 10  1996 Western interconnection blackouts.
However, cascading failure remains a hard problem requiring multiple approaches.
In this paper we pursue another and complementary  approach which is bulk statistical analysis of typical observed cascading data.
One advantage of analyzing real data is that there are no modeling assumptions.

 Bulk statistical analysis can describe the size and extent of cascading from real or simulated cascades
so that blackout risk can be quantified.
For example,  cascading transmission line outages  from utility data \cite{RenCAS08,DobsonPS12} can be used to estimate 
the average propagation of transmission line outages during the cascade as well as the probability distribution of cascading size in terms of number of transmission line outages. Similar efforts for simulated cascades in \cite{KimSJ12,QiPS13} characterize the distribution of blackout size in terms of load shed. These studies quantify the average growth in blackout size during the cascade and the probability  of the cascade growing to a given blackout size.
Observed WECC data for the causes and frequencies of common mode and dependent outages are analyzed in \cite{KeelPESGM12}.

There has also been general progress in describing how cascades spread in simulated cascades,
especially those assuming that cascades propagate by line outages causing static overloads.
Overloaded line cascading outage interactions are analyzed using network resistance distance 
and line outage distribution factors in  \cite{SoltanArxiv14}. Line outage distribution factor calculations of the effect of double contingencies in overloading other lines are used to find all the critical N-2 contingencies in \cite{KaplunovichPESGM13}.
Progression of cascading through sets of lines is described in the simulation of \cite{RahnamayPS14}.
Interactions between cascading line outages are described by line interactions graphs different than the 
transmission network topology in \cite{HinesHICSS13,QiPS15}. 

This paper processes observed transmission line outage data to 
obtain the statistics of how cascades spread in a real power network. 
This is, to our knowledge, the first statistical study of typical cascade spreading and spatial extent based on real data.
The statistics of the manner and extent of real cascade spreading is basic information 
that can support the analysis and mitigation of cascading.
For example, the chance of a cascade spreading a certain amount can inform the design of 
 area monitoring and control to mitigate  cascading, and  
the fraction of cascading interactions at a given distance in the network is of interest in distinguishing 
the mechanisms of cascading that more frequently arise in practice.

  Moreover, there is a large variety of many different simulation models of cascading that are claimed to represent 
 cascading in power networks \cite{TaskForce08,TaskForce11}, and the extent to which the statistics of cascade spreading match the observed statistics 
 serves either to validate the simulation or to suggest ways to improve or disprove the simulation model \cite{WorkingGroup15}.
 It is especially important to make this comparison since the real data incorporates all the mechanisms 
 of cascading whereas the current simulations  only represent a limited and varying subset 
 of  the dozens of plausible cascading mechanisms \cite{TaskForce08,WorkingGroup15}.  More generally, the objective of the validation of the models and simulations with real data is to determine which  mechanisms need to be 
 represented and in what detail in order to be able to do cascading failure risk analysis with 
 confidence in the results.
 To achieve this objective, it is necessary to 
 develop methods of data processing so that 
 the statistics of typical real  cascades can be obtained and compared with the statistics of  simulated cascades. 
 
 Many countries, including the United States and Canada,  collect useful transmission line outage data, and it would seem straightforward to 
 sort this data into individual cascades and determine where on the  network the successive line outages 
 are located, and hence obtain the statistics of cascade spreading. 
 However, this is difficult unless a network model consistent with the outage data is available.
 Indeed, an initial effort using observed North American  transmission line outage data
 encountered 
 substantial difficulties in automating the location of the line outages in a network model that was not consistent with the outage data  \cite{DingMS13}.
 For example, single buses representing a single substation in the observed line outage data can correspond to multiple buses of a detailed 
 network model, and the details of corresponding bus names can differ.
 Single lines in the observed line outage data can correspond to several sections of lines in the detailed network  model.
 Devices such as transformers in the detailed network model need to be accounted for, 
  and the areas and voltage levels covered by the observed line outage data and the detailed network model need to be coordinated.
 Overall, an automated analysis is difficult since the observed line outage data corresponds to a particular reduction of a detailed network model, and it is not 
 straightforward to perform that reduction in order to be able to relate the network  implicit  in the observed data 
 with the detailed network model.
 These difficulties can be overcome to some extent by a sustained combination of automatic and hand processing;
 indeed \cite{DingMS13} processes one year of line outage data for higher voltage levels,
 but it remains challenging and arduous to process enough data for statistically meaningful results.
 These difficulties are not surprising; they are an example of the general difficulty of coordinating different data bases containing different power system network descriptions. 
 This paper describes a much better  approach: we discovered that it is practical to form a satisfactory  network
 directly from the line outage data as explained in section \ref{forming}.

The goal of this paper is to analyze observed cascading data to quantify how
cascades typically spread.
We describe a practical method to process standard utility data to locate outages on a network and obtain some bulk statistics of the spread, and illustrate 
the new method with real  data that is publicly available. Similar data is produced by North American utilities for reporting to NERC, and also by 
some utilities worldwide, so that the method can be applied broadly to existing utility data.

\section{Forming the network from utility data}
\label{forming}

The required data is a list of recorded transmission line outages\footnote{
The analysis could be extended to incorporate other outages such as generators and transformers, but 
since we do not have enough of this data, and the transmission lines capture  the spreading, 
automatic transmission line outages suffice for a first analysis.
}, including the outage start time (to the nearest minute suffices) and the 
names of the buses at both ends of the line, and, for multiple circuits between the same two buses, the circuit number. 
The automatic line outages should be identified, since the
cascade analysis should primarily address the automatic outages. For some purposes it is also useful to know the line length, nominal voltage rating and district.
All this data is standard.  For example, this data is reported by North American utilities in NERC's Transmission Availability Data System (TADS) \cite{BianPESGM14,NERCreport15} and is also collected 
in several other countries.

The transmission line outage data used in this paper starts from 44\,593 automatic and planned line outages\footnote{ Lines that are normally out are ignored.}
recorded by a North American utility
over a period of 14 years starting in January 1999 \cite{BPAwebsite}.
The data requires some cleaning adjustments before the main processing.
Outages in districts remote from the main network, outages of 9 lines rated below 68 kV, outages of 7 lines that did not have bus names for each end of the line
and 10 rural lines that seemed disconnected from the 
main network were all deleted.
About 20  bus names were adjusted to eliminate duplicate forms of the same bus name or to combine buses in the same or adjacent substation.
This left 42\,561 automatic and planned line outages from the main connected network of the utility, with each 
 line outage having a sending end and receiving end bus identified from a list of unique bus 
 names.\footnote{We do not process outages of sections of lines or taps of lines or feeders in forming the network and analyzing the outages.}
 
Then the network model was constructed simply by joining two buses with a transmission line if there was 
in the data an automatic or planned  outage of the line joining those buses.
This procedure produces a subset of the actual network, capturing only those branches that have experienced an outage within the time horizon of the data.
Since it is not obvious how much outage data is needed to form a sufficiently complete network model in this way,
we address and confirm the completeness of the network model formed in this way in section \ref{completeness}.

\begin{figure}[h]
  \centering
 \includegraphics[width=0.9\columnwidth]{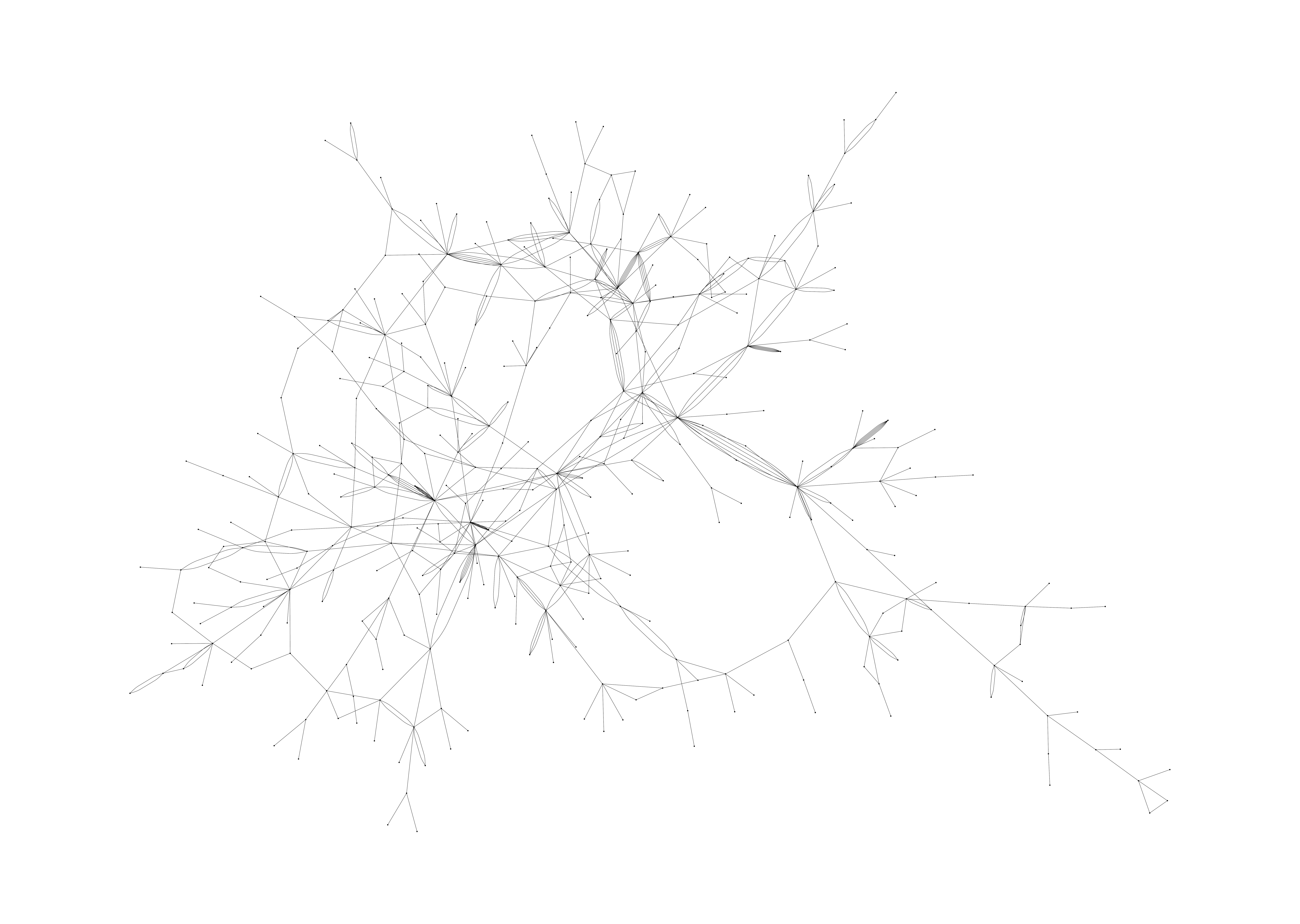} 
  \caption{Network formed from  line outage data. Layout is not geographic.}
  \label{networknolabel}
\end{figure}

\begin{figure}[h]
  \centering
\includegraphics[width=\columnwidth]{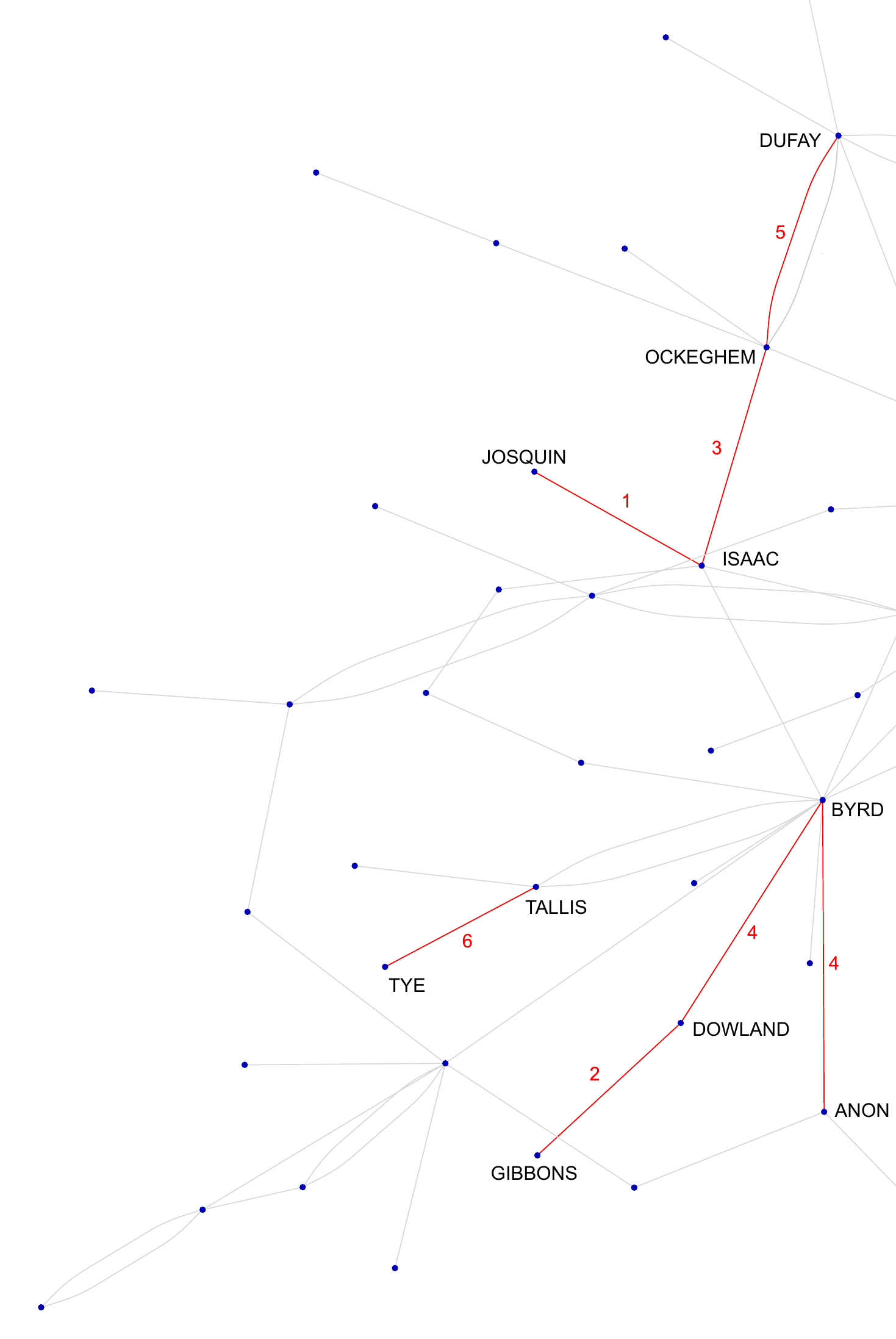} 
\sffamily
  \begin{tabular}{ccc}
  \renewcommand{\familydefault}{\sfdefault}
\scriptsize &\scriptsize outage start time &\\[-3pt]
\scriptsize transmission line&\scriptsize hour:minute&\scriptsize generation \\\hline
\scriptsize JOSQUIN\,$-$\,ISAAC&\scriptsize  15:22 & \scriptsize 1 \\[-1pt]
\scriptsize GIBBONS\,$-$\,DOWLAND&\scriptsize  15:25 & \scriptsize 2 \\[-1pt]
\scriptsize ISAAC\,$-$\,OCKEGHEM&\scriptsize  15:27 & \scriptsize 3 \\[-1pt]
\scriptsize DOWLAND\,$-$\,BYRD&\scriptsize  15:37 & \scriptsize 4 \\[-3pt]
\scriptsize ANON\,$-$\,BYRD&\scriptsize  15:37 & \scriptsize 4 \\[-1pt]
\scriptsize OCKEGHEM\,$-$\,DUFAY No 1&\scriptsize  15:49 & \scriptsize 5 \\[-1pt]
\scriptsize TYE\,$-$\,TALLIS&\scriptsize  15:57 & \scriptsize 6 \\
 \end{tabular}
   \caption{Illustrative example of a cascade of line outages located on the network. The darker and red network lines are the lines that outage. The numbers are the generation number 
 of the outage  and show the order in which the outages occur. Outages occurring in sufficiently quick succession are in the same generation.
The bus names and outage times are changed for the illustrative presentation in this figure. Layout is not geographic. }
  \label{networkcascade}
\end{figure}

The network model obtained from the data is shown in Fig.~\ref{networknolabel}. It is a connected network with 361 buses and 614 lines.
An important practical advantage of forming the network model directly from the outage data is that 
there is then no difficulty establishing the correspondence between the  network model and the outage data; the correspondence is immediate by construction.
For example, an observed cascade obtained from the data set is located on the network as shown in Fig.~\ref{networkcascade}. Fig.~\ref{networkcascade} changes or  omits  identifying details since it is bad practice to publish these when it is not absolutely necessary.

\section{Grouping outages into cascades and generations}
\label{grouping}

Having formed the network model from both the automatic and planned line outages, the analysis of the cascade spreading 
proceeds with only the automatic outages. There are 10\,942 automatic outages in the data.
One motivation for analyzing only the automatic outages is that cascading focusses on uncontrolled outages; for example, NERC defines cascading as ``the
uncontrolled successive loss of system elements triggered by
an incident at any location \cite{NERCweb}."

The structure of cascading is that each cascade starts with initial outages in the first generation 
followed by further outages grouped into generations 2, 3, 4, ... until the cascade stops.
The first step in processing the line outages is to group the line outages  into individual cascades, and then within each cascade 
to group the outages that occur in close succession into generations.
The grouping of the outages into cascades and generations within each cascade is done based on the outage start times 
according to the method of  \cite{DobsonPS12}.
We summarize the procedure  here and refer to  \cite{DobsonPS12} for the details.
The grouping is done by looking at the gaps in start time between successive outages.
If successive outages have a gap of one hour or more, then the outage after the gap starts a new cascade
(note that operator actions are usually completed within one hour).
Within each cascade, if successive outages have a gap of more than one minute, then the outage after the gap 
starts a new generation of the cascade
(note that fast transients and protection actions such as  auto-reclosing  are completed within one minute).
Note that since the outage times are only known 
to the nearest minute, the  order of outages within a generation often cannot be determined.

This procedure applied to the 10\,942 automatic outages yields 6687 cascades.
84\% of these cascades have only one generation of outages and do not spread further.

\section{Network distances}

To quantify the spatial spreading of the cascading line outages, we  specify two
measures of distance in the network between two lines.

The network distance between lines $L_i$ and $ L_j $ in terms of number of buses\footnote{It is common to define the network distance between buses as the minimum number of  lines 
in a path between the buses, and $d^{\rm bus}(L_i,L_j)$ can be conveniently 
evaluated using this network distance between buses:  Except for the case of the distance of a line to itself,  $d^{\rm bus}(L_i,L_j)$ is one plus the minimum bus distance between either of the end buses in $L_i$ 
and either of the end buses in $L_j$. This follows since a path between the midpoints of $L_i$ and $L_j$
with the minimum number of buses must include a path between the end buses of $L_i$ and the end buses of $L_j$ with the 
minimum number of buses.}\footnote{The distance $d^{\rm bus}(L_i,L_j)$ is the same as the network distance between $L_i$ and $L_j$ in the line graph of the network.} is defined as 
\begin{align}
d^{\rm bus}(L_i,L_j)=&\mbox { minimum number of buses in a network path}\notag\\[-3pt]
&\mbox{ joining  midpoint of } L_i  
\mbox{ to midpoint of } L_j.\notag
\end{align}
For example, the distance $d^{\rm bus}$ of line to itself is zero and the distance $d^{\rm bus}$ of a line to a neighboring line with at least one bus in  
common is one.

The network distance between lines $L_i$ and $ L_j $ in terms of miles of transmission line\footnote{$d^{\rm miles}(L_i,L_j)$ can also be conveniently 
evaluated using a network distance between buses:  Except for the case of the distance of a line to itself,  $d^{\rm miles}(L_i,L_j)$ is 
half  the length of $L_i$ plus half the length of $L_j$ plus the minimum bus distance in miles between either of the end buses of $L_i$ 
and either of the  end buses of $L_j$.} 
is defined as 
\begin{align}
d^{\rm mile}(L_i,L_j)=&\mbox { minimum length in miles of a network path}\notag\\[-3pt]
&\mbox{ joining  midpoint of } L_i  
\mbox{ to midpoint of } L_j.\notag
\end{align}

Cascading lines occur in generations and 
we define the network distance between two generations of lines.
(Note that from the point of view of the processing that groups outages into generations,  lines outaging in the same generation outage simultaneously and their outage times cannot be distinguished.)
We write $d$ for the network distance which can either be
 in terms of number of buses or   in terms of miles.
Then the mean network distance between generation of lines $G_i$ and generation of lines $ G_j $ is defined as 
\begin{align}
d_{\rm mean}(G_i,G_j)&=\mbox{ mean}\{d( L_i ,L_j), \ L_i \mbox{ in } G_i \mbox{ and }  L_j \mbox{ in } G_j \}\notag
\end{align}
and the 
maximum network distance between generation $G_i$ and generation $ G_j $ is defined as 
\begin{align}
d_{\rm max}(G_i,G_j)=&\mbox{ max}\{d( L_i ,L_j), \ L_i \mbox{ in } G_i \mbox{ and }  L_j \mbox{ in } G_j \}.\notag
\end{align}

\section{Statistics of cascade spreading}

\begin{figure}[h]
  \centering
 \includegraphics[width=\columnwidth]{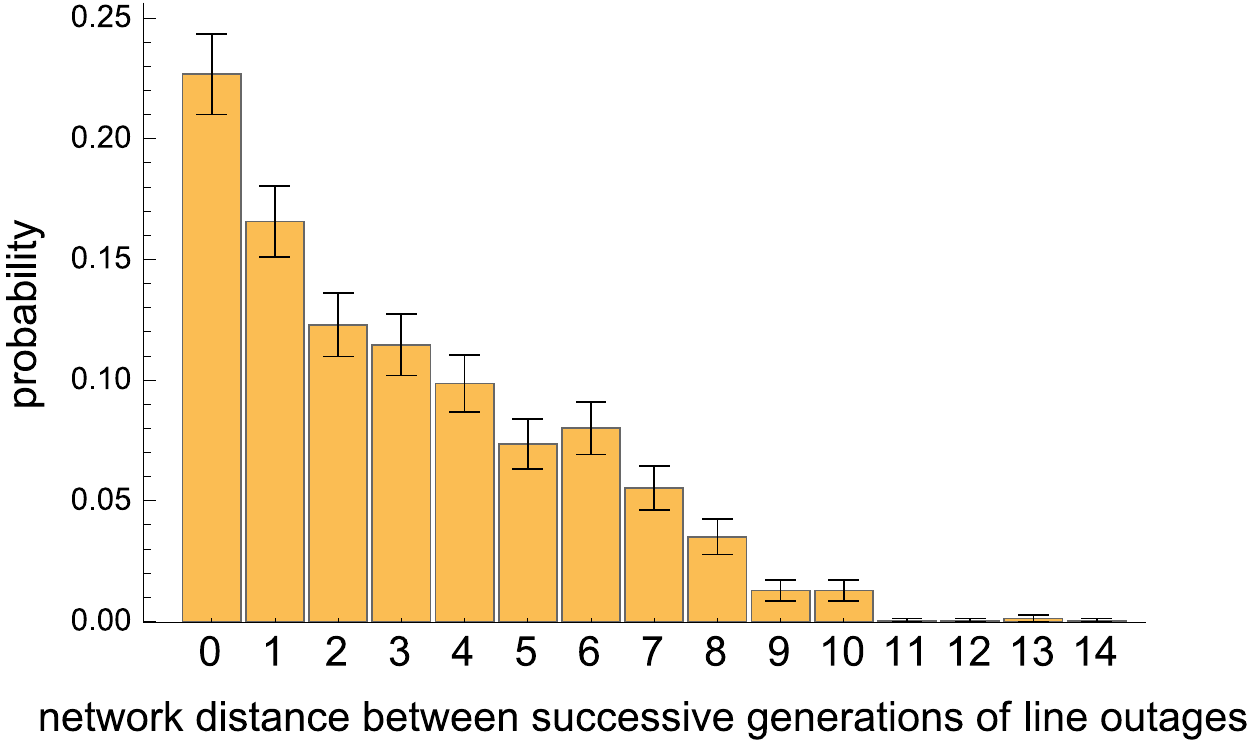} 
  \caption{Probability distribution of  network distance $d_{\rm mean}^{\rm bus}( G_i ,G_{i+1})$ between successive generations of line outages. The error bars show a 95\% confidence interval. }
  \label{BPACYgenhops}
\end{figure}
\begin{table}[h]
\centering \caption{Probability distributions of  network distances}
\label{tableBPACYgenhops}
\centering
\begin{tabular}{c|cc|c}
&\multicolumn{2}{|c|}{$d_{\rm mean}^{\rm bus}$}&$d_{\rm maxspread}^{\rm bus}$\\
&\multicolumn{2}{|c|}{ Between successive generations}&Max from initial\\
&&probability given\\
distance&probability&distance\,$>$\,0&probability\\\hline
0&0.227$\pm$0.017&&0.777$\pm$0.010\\1&0.166$\pm$0.015&0.214$\pm$0.019&0.088$\pm$0.007\\2&0.123$\pm$0.013&0.159$\pm$0.017&0.023$\pm$0.004\\3&0.115$\pm$0.013&0.148$\pm$0.016&0.023$\pm$0.004\\4&0.099$\pm$0.012&0.127$\pm$0.015&0.018$\pm$0.003\\5&0.073$\pm$0.010&0.095$\pm$0.013&0.016$\pm$0.003\\6&0.080$\pm$0.011&0.103$\pm$0.014&0.020$\pm$0.003\\7&0.055$\pm$0.009&0.071$\pm$0.012&0.015$\pm$0.003\\8&0.035$\pm$0.007&0.045$\pm$0.009&0.011$\pm$0.003\\9&0.013$\pm$0.004&0.017$\pm$0.006&0.004$\pm$0.002\\10&0.013$\pm$0.004&0.017$\pm$0.006&0.003$\pm$0.001\\
\hline
\omit&\multicolumn{3}{c}{$\pm$ errors are 95\% confidence intervals}\\
 \end{tabular}\end{table}
\begin{figure}[h]
  \centering
 \includegraphics[width=\columnwidth]{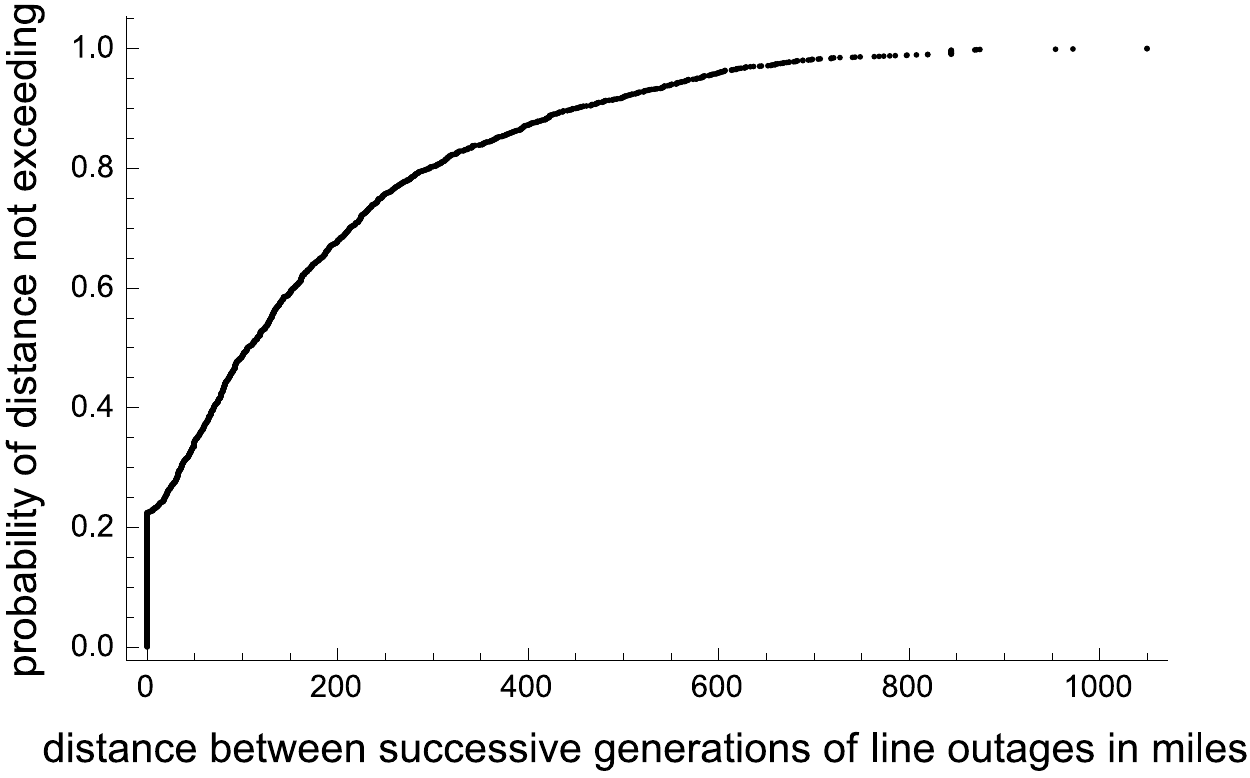} 
  \caption{Cumulative probability distribution of the network distance $d_{\rm mean}^{\rm mile}( G_i ,G_{i+1})$ between successive generations of line outages. }
  \label{BPACYcdfdistance}
\end{figure}
There are 6687 cascades and 
84\% of these cascades have only one generation and so do not spread further. 
To analyze the spreading cascades, we exclude the cascades with only one generation, and only analyze the 
remaining 1098  cascades with more than one generation.

We are interested in the distance between successive generations of line outages
$
d_{\rm mean}( G_i ,G_{i+1})
$
where  $G_i$ and $G_{i+1}$ are successive generations in the same cascade.
There are 2426 such pairs of successive generations in the data. (There are no successive generations in 
the 5589  cascades with only one generation.) 

 The statistics of the network distance for these  successive generations in terms of number of buses is shown in Fig.~\ref{BPACYgenhops} and  Table~\ref{tableBPACYgenhops}.
The mean number of buses between successive generations is 2.9 and the median number of buses between successive generations  is 2.
The most frequent number of buses (23\%) is zero; these are cases in which an outaged line is restored but outages again after more than one minute and less than one hour after the initial outage.
Next most frequent (17\%) is one bus, in which a neighboring line outages. But more than one bus, or equivalently not a repeat outage and not a neighbor, 
has frequency 60\%. So, while neighboring lines do outage, this is less likely than the same line tripping again and much less likely than a non-neighboring line outaging.

The distribution of the distance in network miles of successive generations of the spreading cascades is shown in Fig.~\ref{BPACYcdfdistance}. 
Fig.~\ref{BPACYcdfdistance} shows that 
half of the successive generations spread more than 100 miles, one third of the successive generations spread more than 200 miles, and one eighth  of the successive generations spread more than 400 miles.

In the processing described so far, we have counted a repeated outage of the same line after more than  one minute 
delay as one additional outage, and the distance spread in this case is zero. While this is reasonable since  repeated outages have more impact than a single unrepeated outage, one could alternatively regard the repeat outages as the same as the original outage and count them only once. Then all the successive outages in a cascade move in the network a distance that is greater than zero.
The effect of this alternative assumption on the spreading statistics is obtained by conditioning the 
probabilities on the spreading distance being greater than zero and the results for the bus network distance are also shown in 
Table~\ref{tableBPACYgenhops}.

\begin{figure}[h]
  \centering
 \includegraphics[width=\columnwidth]{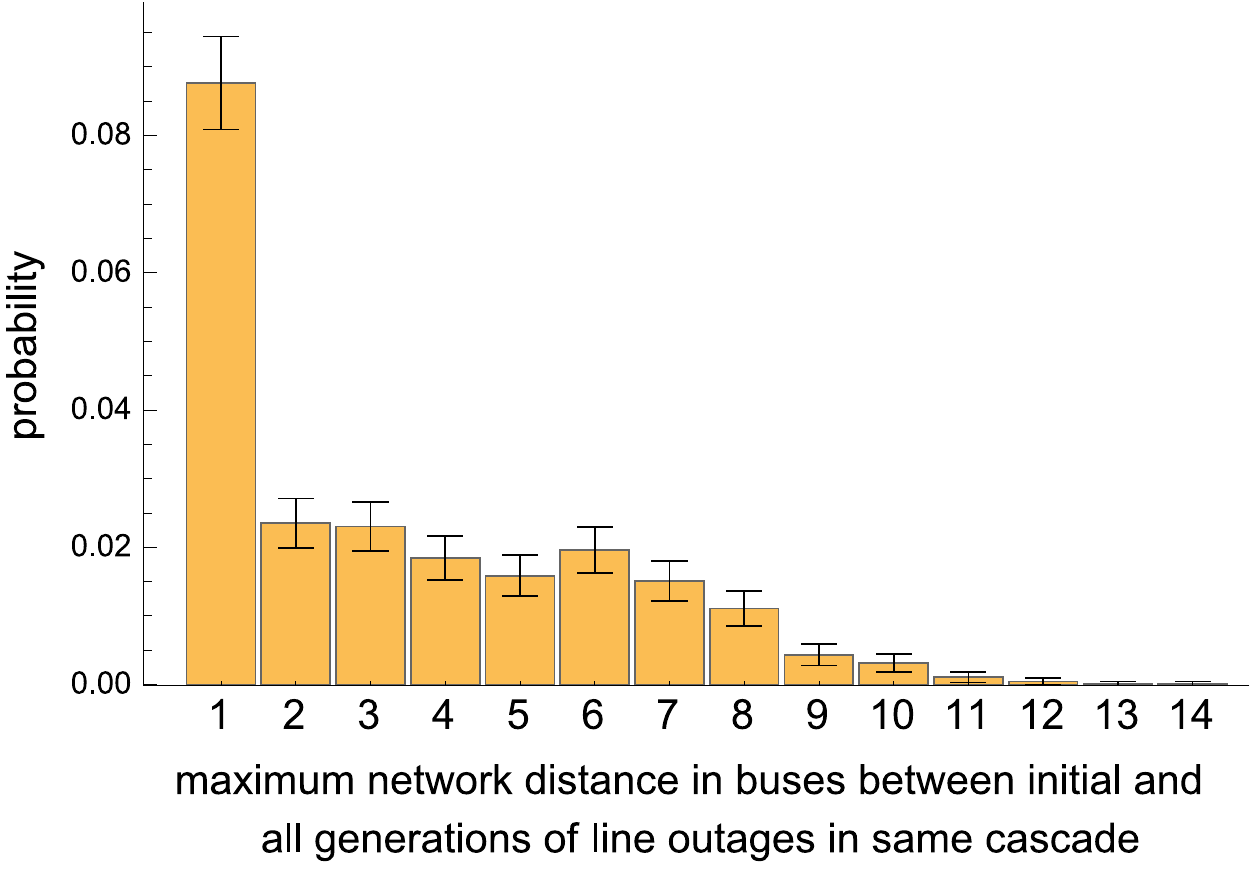} 
  \caption{Probability distribution of the network distance $d_{\rm maxspread}^{\rm bus}$ which is the maximum network bus distance between the initial 
  generation of line outages and any generation of line outages in the same cascade. The error bars show a 95\% confidence interval. Only $d_{\rm maxspread}^{\rm bus}>0$ is plotted.
  The probability of a cascade not spreading ($d_{\rm maxspread}^{\rm bus}=0$) is 0.78 with 95\% confidence interval $\pm0.005$.}
  \label{BPACYgenmaxhops}
\end{figure}

\begin{figure}[h]
  \centering
 \includegraphics[width=\columnwidth]{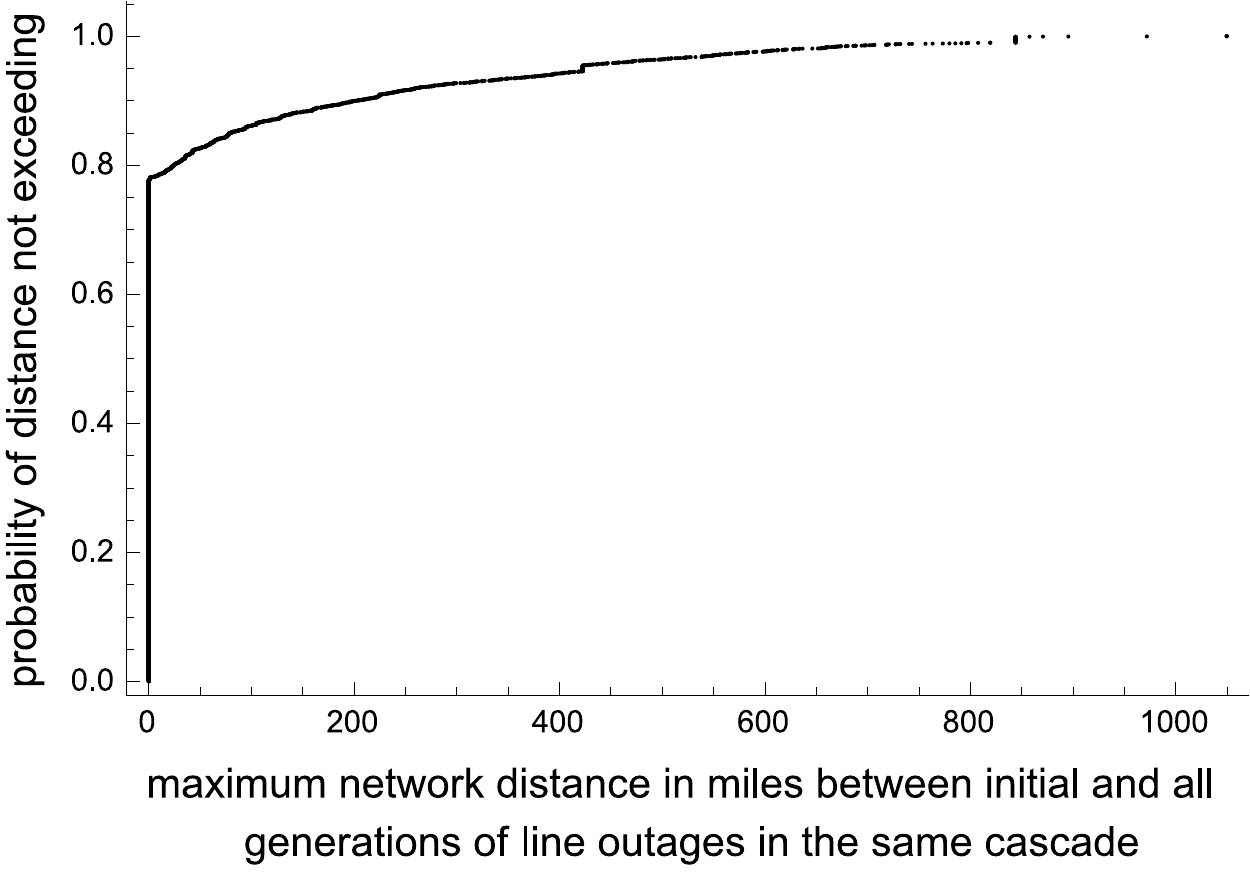} 
  \caption{Probability distribution of the network distance $d_{\rm maxspread}^{\rm mile}$ which is the maximum distance in miles between the initial 
  generation of line outages and any generation of line outages in the same cascade.}
  \label{BPACYcdfmaxdistance}
\end{figure}

We are also interested in the maximum distance that a cascade spreads from the initial generation of outages $G_0$:
\begin{align}
d_{\rm maxspread}(C_k)&=\mbox{max}\{d_{\rm max}( G_0 ,G_i),   \notag\\ &G_0 \mbox{ the initial generation in cascade $C_k$ and }  \notag\\ &G_i \mbox{ any generation in cascade $C_k$}\}\notag
\end{align}
This maximum spreading distance  is shown in Fig.~\ref{BPACYgenmaxhops} and Table~\ref{tableBPACYgenhops} for number of buses  and Fig.~\ref{BPACYcdfmaxdistance} for network miles.
Most of the probability of zero spreading is due to cascades with only one line outage. If the cascades with only one  line outage are excluded, the mean maximum distance spread is 3.8 buses.

All the spreading statistics show the effects of the finite size and edges of the network. 
In terms of network bus distance, the diameter of a network is the maximum possible distance $d^{\rm bus}(L_i,L_j)$ 
between any two lines $L_i$ and $L_j$.
The diameter of the network is 15, so this 
is an upper bound to the spreading results shown in Figs. \ref{BPACYgenhops}, \ref{BPACYgenmaxhops},  and Table~\ref{tableBPACYgenhops}.
There are two types of cascades: cascades that are confined to the network lines for which the data includes all the outages, and cascades that 
involve lines outside the network that have missing data.
The cascades that involve lines outside the network have spreads that can exceed 
the spread confined to the network and affect the results in Fig. \ref{BPACYgenmaxhops} by tending to increase moderate  spreads,
tending to reduce larger observed spreads such as spreads of more than 10 buses, and eliminating cascades of more than 15 buses.
This ``network edge effect" is of interest for future work in trying to quantify 
how many cascades spread to or from a neighboring power system area.

\section{Independent outages and dependent  outages}

Cascading arises from a variety of types of dependent outages, and it is useful to check that independently occurring  outages do not contribute much to the results.
Independent outages occur randomly throughout the year.
Most of these independent line outages are isolated in time from each other and from dependent outages, and do not 
contribute to the measures of cascading effects used in this paper.
By  chance, occasionally these independent line outages occur in close proximity to each other or to dependent outages in time and do contribute to the measures of  cascading.
This section quantifies the contribution of statistically independent outages towards the measures of  cascading used in this paper.

Each cascade contains at least one initiating outage, and we  assume that the first initiating outage\footnote{If there are several initiating outages at the same time, then we arbitrarily choose one of these to be the first one.} in each cascade is independent, 
 and that a fraction $R$ of the remaining outages in all the cascades are also independent outages. 
 Then, since there are $6687$ initiating outages and $4255$ remaining outages, there are $6687+4255R$ independent outages in the $10\,942$ observed outages, 
and this corresponds to an independent outage rate of $0.055+0.035R$ per hour.
We model the independent outages as a Poisson process \cite{CarrerasCAS04} with this rate.
In particular, the time differences between independent outages are exponentially distributed with rate parameter $0.055+0.035R$.

It is convenient to call the outages  that remain after the first initiating outage of each cascade is omitted ``remaining outages."
A remaining independent outage is processed as belonging to a cascade when it occurs after the  initiating outage but no later than one hour after the last outage of a cascade. 
 The average time between the  initiating outage and last outage of a cascade is 7 minutes or 0.12 hour.
 Therefore, on average, 
 a remaining independent outage is processed as belonging to a cascade when it occurs less than 1.12 hour after the initiating outage of a cascade. 
Therefore, assuming that the preceding independent outage is the initiating 
outage\footnote{
In the much rarer case that the preceding independent outage during the cascade is not the initiating outage, the approximation (\ref{reqn}) is not much different since in this case the 
remaining independent outage is processed as belonging to a cascade when it occurs on average less than a time $T$ after the 
preceding outage, where $1.0<T<1.12$.},
the fraction $R$ of  remaining independent outages that are processed as belonging to a cascade is
\begin{align}
R&= P[\mbox{time difference with preceding outage $<$ 1.12 hour}]\notag\\
&= 1-\exp[-(0.055+0.035R)1.12]
\label{reqn}
\end{align}
Solving (\ref{reqn}) numerically gives $R=0.06$. That is, approximately 6\% of the remaining outages, or 4\% of all outages,
are independent but classified as cascading outages.

One way to appreciate the strong effect of cascading dependence in the cascade spreading results is to 
remove the dependence by retaining the observed outages, but 
assigning them artificial and random outage times sampled from a Poisson process.
Then, with the same processing described in section~\ref{grouping}, the  number of cascades increases from 6687 to 9956 because there are more cascades with only one outage, but 
these initial outages propagate very weakly as shown in Table~\ref{table_generations}; the average propagation of outages\footnote{Details of the average propagation definition are in \cite{RenCAS08}.} reduces from 0.28 to 0.08.

\begin{table}[h!]
\centering \caption{Number of line outages in generations 0 to 10}
\label{table_generations}
\centering
\begin{tabular}{@{\extracolsep{-2.pt}}
ccc ccc ccc cc}
\multicolumn{11}{c}{generation number}\\
$0$&$1$&$2$&$3$&$4$&$5$&$6$&$7$&$8$&$9$&$10$\\
\hline\\[-6pt]
\multicolumn{11}{c}{outages with cascading dependence; average propagation = 0.28}\\
7911&1347&497&272&170&114&90&63&54&40&40
\\[2pt]
\multicolumn{11}{c}{~outages with artificial, random times; average propagation = 0.08}\\
9975&841&69&3&1&0&0&0&0&0&0
\\[2pt]
\hline
\end{tabular}
\end{table}

While it is useful for some purposes, such as classifying outages and their mechanisms,  to find out how much independent outages contribute to 
the cascading results by being lumped together with dependent outages, it should also be emphasized 
that the power system operators have to deal with multiple outages closely spaced in time regardless
of their independence or dependence.

One consequence of our analysis method of grouping the outages into generations is that it classifies automatic outages as initial outages (in the first generation),
or as dependent outages (in second or higher generations). 
Of the 10\,942 automatic outages there are 7911 initial outages  (comprising 72\%) and
3031 dependent outages  (comprising 28\%).

Since many of the mechanisms for initial outages 
differ from the mechanisms for dependent outages, it can be expected that there can be some differences between the initial and dependent outages, as observed in simulated cascades \cite{CarrerasHICSS12}.
We examine the most frequently involved lines in initial outages and in dependent outages.
One half of the 20 lines most frequently involved in initial outages differ from the 20 lines most frequently involved in dependent outages. And one third of the 100 lines most frequently involved in initial outages differ from the 100 lines most frequently involved in dependent outages.

It is also very useful to analyze the causes of the initial outages from utility data since some of these causes can be mitigated,
and mitigating the initial outages is one way to reduce cascading outages \cite{BianPESGM14,NERCreport15}. 
This paper does not address this useful aspect of cascade analysis and mitigation because this paper has the 
complementary objective of opening up possibilities for analysis and mitigation of the dependent cascading outages that 
follow the initial outages.

 \begin{figure*}
\centering
\begin{minipage}{.48\textwidth}
  \centering
  \includegraphics[width=\textwidth]{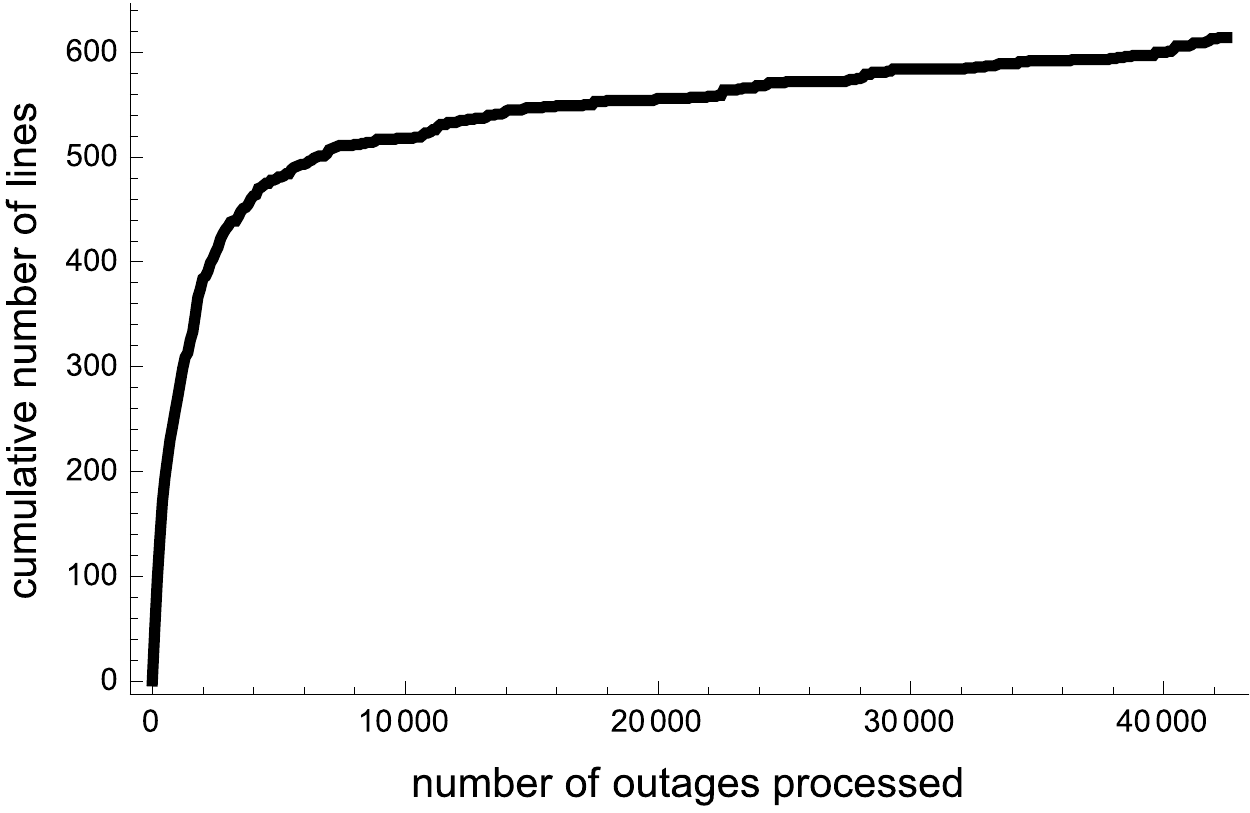}
  \caption{Cumulative number of lines in network as the number of outages processed increases.
}
  \label{BPACYcountcumlines}
\end{minipage}%
~
\begin{minipage}{.48\textwidth}
  \centering
  \includegraphics[width=\textwidth]{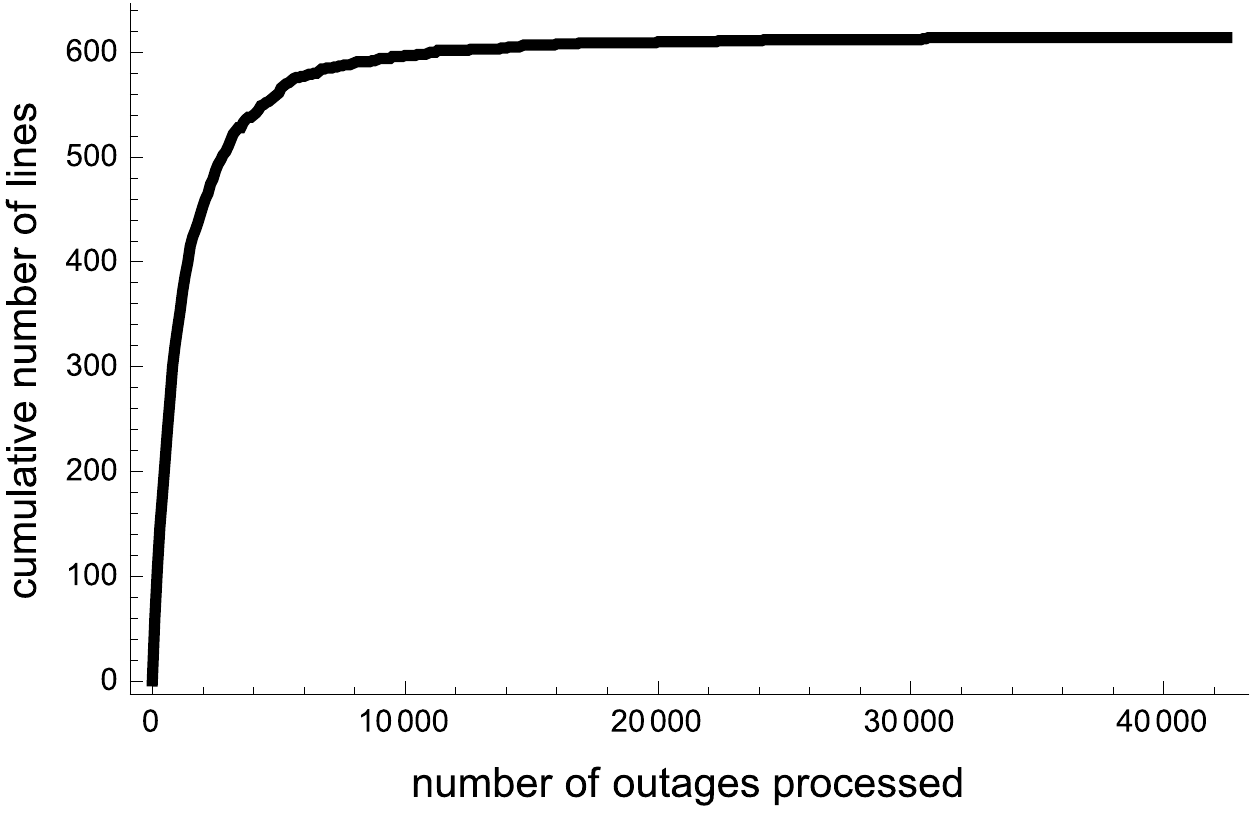}
  \caption{Cumulative number of lines in network as the number of reordered outages processed increases.}
  \label{BPACYcountcumlinesreordered}
\end{minipage}
\vspace{-10pt}
\end{figure*}

\section{Completeness of the network}
\label{completeness}

As more outages are processed, outages of new lines are encountered, and the network formed from all the outages processed so far becomes more complete,
and, if sufficiently many outages from a fixed network are processed, the network  formed from all the outages processed so far  converges to the entire network.
This section examines this convergence to show that enough outages were processed to form a good approximation of the 
network. This verifies building the network from the outage data.

In practice, the network slowly changes as new lines are added and old lines retire. 
Therefore the network formed from the outages includes both 
 retired and new lines over the time period of the processed outages.

 Both convergence towards a complete network and the effect of the network changing can be seen in Fig.~\ref{BPACYcountcumlines}.
Fig.~\ref{BPACYcountcumlines}  initially shows the number of lines in the network converging as more outages are processed and then 
finally increasing slowly as new lines are added in the later outages.  To confirm the convergence, we want to remove the effect of the network changing from 
Fig.~\ref{BPACYcountcumlines}. This is done by splitting the outage data into two halves at the midpoint, reversing the order of the 
data in the second half, and then interleaving the reversed second half with the first half.
To give a small example, if 10 outages were originally in the order 1,2,3,4,5,6,7,8,9,10 then the reordering yields 1,10,2,9,3,8,4,7,5,6.
The problem of confirming convergence arises from new lines added in the later, converging portion of the data 
(new lines added in the earlier portion of the data only affect the transient before the convergence),
and the reversal of the second half of data ensures that these new lines are likely to appear in the data before 
convergence.
(Note that simply reversing all the data does not work: the new lines added during the converging portion of the data 
would now appear before convergence, solving the problem of the new lines, but now there would be a new problem of 
lines retiring at the beginning of the data appearing as added lines at the end of the reversed data.)

The reordered outages that remove the effect of the changing network from the convergence are shown
 in Fig.~\ref{BPACYcountcumlinesreordered} and the convergence is clear.
The  network constructed from the data would ideally include all the lines that have been present for a portion of or all of the time period,
and the convergence analysis shows that the network constructed with the data converges to almost all such network lines.

\section{Sensitivity to cascade interval}

We check the sensitivity of the results to the one-hour minimum interval between 
cascades assumed in processing the data in section \ref{grouping}.
Changing the minimum interval between cascades from one hour to 30 minutes 
causes the number of cascades to increase from 6687 to 7332 and changes the results in Table~\ref{tableBPACYgenhops}
to the results shown in Table~\ref{tableBPACYgenhops30}. The results are close; probabilities change by less than 0.01, with the exception of the 
lowest distance results in each column, which change by less than 0.04.
\begin{table}[h]
\centering \caption{Probability distributions of  network distances\newline with cascade interval 30 minutes}
\label{tableBPACYgenhops30}
\centering
\begin{tabular}{c|cc|c}
&\multicolumn{2}{|c|}{$d_{\rm mean}^{\rm bus}$}&$d_{\rm maxspread}^{\rm bus}$\\
&\multicolumn{2}{|c|}{ Between successive generations}&Max from initial\\
&&probability given\\
distance&probability&distance\,$>$\,0&probability\\\hline
0&0.259$\pm$0.020&&0.805$\pm$0.009\\1&0.179$\pm$0.018&0.242$\pm$0.023&0.090$\pm$0.007\\2&0.125$\pm$0.015&0.168$\pm$0.020&0.022$\pm$0.003\\3&0.107$\pm$0.014&0.144$\pm$0.019&0.020$\pm$0.003\\4&0.090$\pm$0.013&0.122$\pm$0.018&0.015$\pm$0.003\\5&0.066$\pm$0.012&0.089$\pm$0.015&0.012$\pm$0.002\\6&0.071$\pm$0.012&0.096$\pm$0.016&0.014$\pm$0.003\\7&0.047$\pm$0.010&0.064$\pm$0.013&0.009$\pm$0.002\\8&0.030$\pm$0.008&0.041$\pm$0.011&0.007$\pm$0.002\\9&0.012$\pm$0.005&0.016$\pm$0.007&0.003$\pm$0.001\\10&0.011$\pm$0.005&0.015$\pm$0.007&0.002$\pm$0.001\\
\hline
\omit&\multicolumn{3}{c}{$\pm$ errors are 95\% confidence intervals}\\
 \end{tabular}\end{table}

\section{Comparing cascade spread statistics from the OPA simulation with the observed data}
\label{OPA}

We make an initial comparison between the statistics of cascade spreading simulated by the OPA simulation
and the statistics of the observed cascade spreading data from the previous sections.
The intent is to show a specific example of using the paper results for improvement and validation of models,
and to show how some technical issues in such a comparison may be addressed.

We start by briefly summarizing the OPA simulation, which includes a fast time scale for cascading transmission line outages 
and a slow time scale for the complex systems feedback shaping the reliability; for details see \cite{CarrerasCH04}. 

The fast time scale of the cascading line outages and blackouts is of the order of minutes to hours. The cascading blackouts are modeled by overloads and outages of lines determined in the context of  a standard
 DC load flow model of the network power flows and 
generator power dispatch optimized by standard linear programming. 
The successive calculations in the simulation naturally  produce generations of line outages in each cascade. If lines outage in a generation, the model recomputes and check the load flow for overloaded lines. The overloaded lines outage probabilistically, and if any of these overloaded lines outage they form the next generation 
of line outages. If none of the overloaded lines in a generation outage, the cascade stops.

The slow time scale of the OPA simulation in which the power system evolves
 is of the order of days to years. In the slow timescale,  the load power demand slowly increases and transmission lines involved in blackouts are upgraded as engineering responses to blackouts and maximum generator power is increased in response to the increasing demand. These slow opposing forces of load increase and network upgrade self organize the system to a complex system dynamic equilibrium that  is close to the critical points of the system \cite{CarrerasCH04,NewmanREL11,DobsonCH07}. The results used here were obtained from OPA in this complex system dynamic equilibrium condition.

OPA was validated on a model of WECC with respect to historically observed statistics in \cite{CarrerasHICSS13}, using a 
1553-bus network model of WECC developed in a California Energy Commission project for analysis of extreme blackout events \cite{MorganCEC11}. The OPA parameters used were derived from WECC data.
The simulated and observed statistics compared were the distribution of blackout sizes, the number of line outages, the number of generations, and the average propagation of number of line outages between generations. 
The simulated and observed statistics agreed well, except for the average propagation of number of line outages in the later cascade stages.
For the present paper, we extend the comparison to the statistics of cascade spatial spreading using the same 1553-bus network and OPA parameters as in \cite{CarrerasHICSS13}.

16\ 788 cascades and 28\ 361 line outages across the entire WECC were simulated with OPA. Many of these cascades occur wholly or partially outside of the Northwest region\footnote{
The Northwest region can be determined as WECC bus numbers in the range 40\ 000 to 49\ 999.} of WECC that covers the collection area for the observed data analyzed in the paper. 
To approximate the conditions of the observed data, we limited the analysis to the 6534 line outages that occurred inside the Northwest region.
That is,  the analyzed simulated outages correspond to the cascades or parts of cascades that occurred inside the Northwest region.
This yielded 6534 outages which are organized into 2768 cascades and 5082 generations using the method of section III.
These were then processed to obtain the distances of  section IV between generations of line outages using the network distance on the 1553-bus network.

\begin{figure}[t]
  \centering
 \includegraphics[width=0.8\columnwidth]{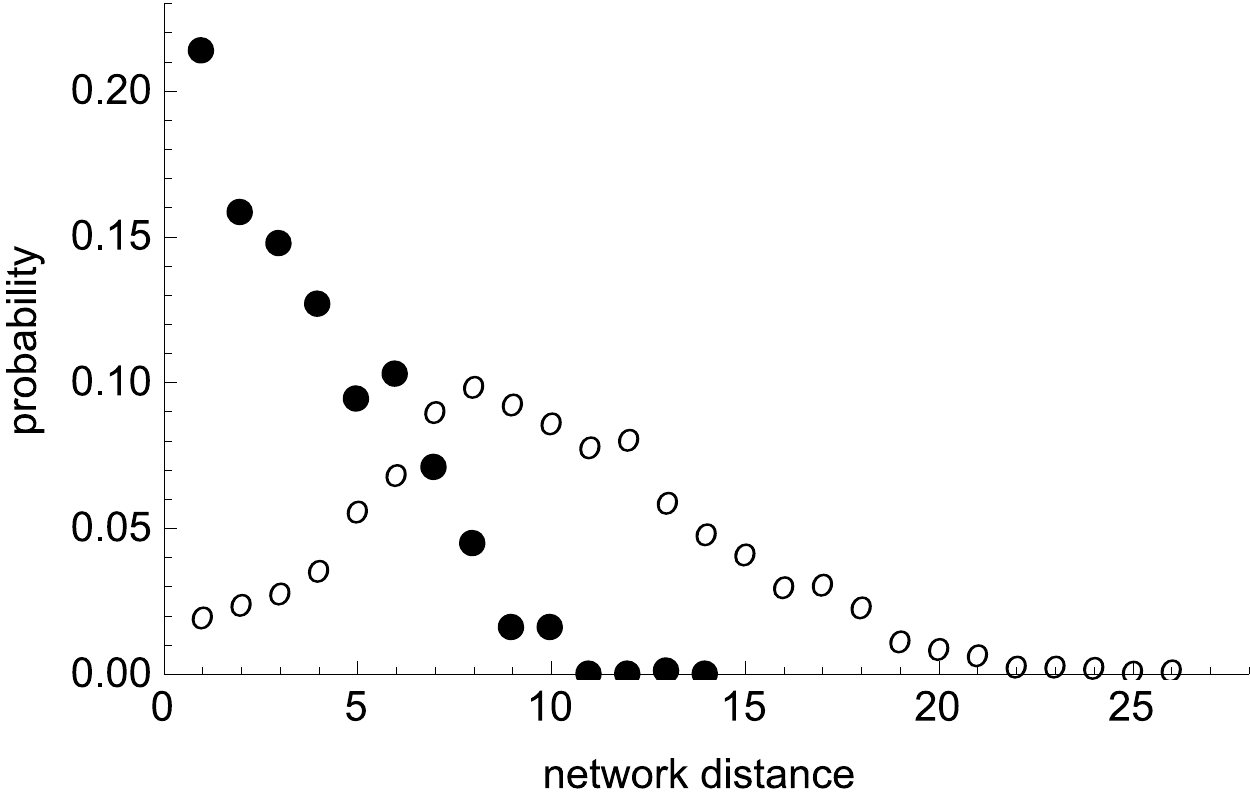} 
  \caption{Probability distributions of  network distance $d_{\rm mean}^{\rm bus}( G_i ,G_{i+1})$ between successive generations of observed line outages on the formed network (dots) and 
  simulated outages on the 1553-bus network (circles).}
  \label{comparehops}
  \vspace{10pt}
  \centering
 \includegraphics[width=0.8\columnwidth]{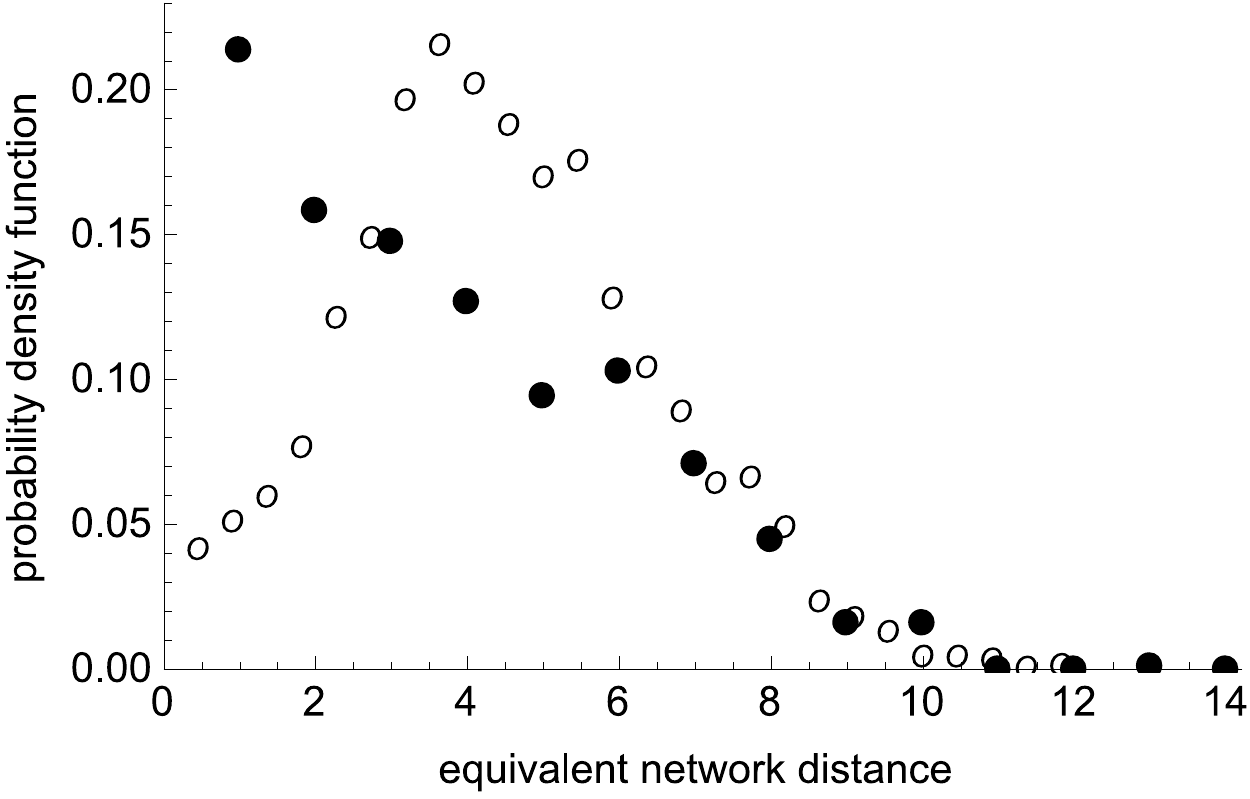} 
  \caption{Probability density functions  of  network distance $d_{\rm mean}^{\rm bus}( G_i ,G_{i+1})$ between successive generations of observed line outages (dots) and simulated line outages  (circles) with the 1553-bus network distances scaled to formed network distances.}
  \label{compareequivalenthops}
\end{figure}

Fig.~\ref{comparehops} shows the probability distribution of distance in the 1553 network between successive generations of simulated outages as open circles.  The mean distance in the 1553 network  between the simulated successive generations of outages is 9.8.
Since OPA (in common with other cascading failure simulations) does not simulate repeated outages of the same line,
the simulated results should be compared with the probability distribution of nonzero distances in the formed network between the observed successive generations of outages. This is the data of the second column of Table \ref{tableBPACYgenhops} and it is plotted as the solid dots in Fig.~\ref{comparehops}.
The mean distance in the formed network between the observed successive generations of outages is 3.8.

A problem with the comparison in Fig.~\ref{comparehops} is that the network distances are measured in different network representations of a similar area of the power system. To correct this, we computed the mean network distance between 10\ 000 pairs of randomly chosen buses in each network, yielding a mean bus distance of 14.0 in the 1553 bus network and a  mean bus  distance of 6.4 in the network formed from the data. The ratio of the bus distances, $6.4/14.0 = 0.46$, is applied as a scaling factor to the simulated distances to allow the comparison with the same distance scale in Fig.~\ref{compareequivalenthops} (to allow direct comparison of the probabilities despite the distance scale change, Fig.~\ref{compareequivalenthops} shows a probability density on the vertical scale). The mean distances expressed in terms of the distance for the formed network are now 4.5 for the simulation and, as before, 3.8 for the observed data.

Fig.~\ref{compareequivalenthops} shows agreement between the statistics of spreading between the simulated and observed data for 
long-range cascading interactions and disagreement for shorter-range interactions.
In particular, the simulated data shows fewer interactions at distances 1 or 2 and more interactions at distances 4 or 5 than the observed 
data. 
Beyond showing which aspects of reality are well described by the simulation and for which purposes the predictions of the simulation are validated,
a particular value of this comparison of spreading statistics is that it suggests the aspects of the simulation to be reconsidered to
improve the match. In this case, the results indicate that the modeling of short range cascading interactions is not captured well enough
by OPA. Obvious candidates for improvement in the short-range modeling would include protection system effects (such as hidden failures \cite{ChenEPES05}) and representing parallel transmission lines in the network.

\section{Conclusions}
It is fundamental to the study of cascading blackouts in transmission networks to be able to  process and characterize real cascading outage data. 
This paper gives new methods to process the spread of transmission line outages from standard utility data already collected by utilities and gives the first statistical characterization of how the 
cascading outages typically spread on the network. We also discuss the opportunities in cascading risk analysis opened up by these 
new methods, including the validation of simulations and models for cascading risk analysis that is needed to advance the field.
The quantification of typical cascade spatial spreading in this paper complements and augments 
 the quantification of cascading propagation in terms of number of line outages in \cite{DobsonPS12}.

\subsection{Contributions to methods of outage data analysis}

The paper contributes new methods of analyzing standard outage data:

\begin{itemize}

\item We solved the problem of obtaining a network model at the same level of detail as and compatible with the outage data by 
using the outage data itself to form the network. This is shown to be an effective and practical solution to an otherwise messy problem coordinating 
different network descriptions.
Then the recorded outages can be readily located on the network so that their spread can be observed in terms of network distance between generations of outages.

\item We demonstrated methods to verify that a substantially complete network is formed from the line outage data.
Even when the network changes over the period of observation, reordering the data can verify the network completion.

\item Cascading is initial outages followed by dependent outages. 
The processing methods include a small  fraction of independent outages among the dependent outages, and we show how to estimate the 
fraction of independent outages.

\item We define metrics describing the  average distances between generations of line outages on the network in terms of 
both average number of buses between the generations and network miles.

\end{itemize}

\subsection{Contributions of the results of the data analysis}
We present for the first time some basic statistics of real cascades spreading on a power transmission network.
The spreading is quantified in terms of the minimum number of buses or network miles  between generations of cascading  line outages.
In the case of the average minimum number of buses between generations, a generation of outages  is followed by 
a repeated outage of the same line in about one quarter of the cases, and only one sixth are followed by a neighboring line outage.
(Even if we ignore the repeated outages,  less than one quarter are followed by a neighboring line outage.) A generation of cascading outages  is followed by 
  an outage of a non-neighboring line in over half of the cases.
As detailed below, these statistics of typical cascade spreading can be used to help 
validate cascading models and simulations,  understand the mechanisms of dependent 
cascading outages, design cascade mitigation schemes, and develop new cascading models.

The data shows that the  lines most involved in initial outages and the subsequent propagation 
of dependent outages differ somewhat, as might be expected from the different mechanisms involved.
Also, there are dramatic differences in the amount of propagation  between 
realistic outages that have cascading dependencies and outages that occur at artificially randomized times.

We show that, after the initial automatic outage, the following cascades of mostly dependent outages  
contain about 6\% independent outages.
The distinction between independent and dependent outages is important in understanding and mitigating cascading, but it is also worth noting that 
in any case
the power system operators have to cope with multiple outages regardless of their cause.

While the data used in the paper is from one large transmission operator, the methods can be applied much more broadly  because 
the data is already routinely collected by some transmission operators internationally.
 In the USA, since TADS data is reported by all transmission operators, the approach can be applied  by any transmission operator or 
 by reliability organizations that aggregate the  TADS data.

\subsection{Using spatial spreading data to validate models and simulations for cascading risk analysis}

A large variety of cascading outage models and simulations have been proposed. For example, a 2008 survey paper references a sample of about 25 different models and simulations and many more have been subsequently proposed. However, there has been very little quantitative validation of these models with real data in the sense of reproducing observed cascade statistics  so that they can 
be relied upon  to quantify cascading risk.\footnote{
In many cases,  simulated cascades are judged to be credible, or a limited selection of the dozens of  mechanisms involved in cascading failure are reasonably approximated. Exceptions where some aspects have been quantitatively validated with observed data include
\cite{KosterevPS96,DobsonPS12,CarrerasHICSS13}.
 For a detailed account of validation approaches and current needs see \cite{WorkingGroup15}.}
 There is a strong need for this validation so that the most important mechanisms of cascading risk analysis 
 can be determined and represented at an appropriate level of detail.
The statistics of cascading spread in this paper are a new contribution to the observed cascade statistics.
This will enable the validation and improvement of cascading models and simulations 
that are close to the reality of power systems and the rejection of models that are unrealistic.

Two examples of this validation process are:

\begin{itemize}
\item Section~\ref{OPA} shows an example of this validation process by comparing the statistics of the 
next generation spreading of line outages in the OPA simulation of cascading with the 
observed statistics, after an appropriate normalization of the distances in the simulated network.
We show how the comparison of spatial spreads distinguishes the matches obtained for long-range and short-range 
 cascading mechanisms, giving insight into which mechanisms need to be differently modeled to get a better match.

\item Our data clearly shows a substantial fraction of non-local propagation of outages.
Therefore cascading models from complex network theory that hypothesize nearest-neighbor propagation on the 
topology of the electrical network are inconsistent with the data observed in this paper.
\end{itemize}

\subsection{Opening up other possibilities in cascading risk analysis}

Other possible research directions based on the cascading spread data include:

\begin{itemize}
\item When analyzing outage data to better understand cascading, dependent outages can now be classified according to local or more global mechanisms
according to how far they are from preceding outages in the cascade.
The larger risk analysis context is that one can start from available outage data and then 
use the interaction distances and other attributes to classify the observations of dependent outages into groups of mechanisms.\footnote{This general ``top-down" and data-driven approach is complementary to detailed modeling of a particular dependent outage mechanism 
and then seeking data for the detailed model. }
\item 
The observed statistical data on cascading spread could enable approximate high-level stochastic models of the effect of cascading. For example, given some initial damaged components  outaging in an earthquake \cite{RomeroES15,RomeroPS13}, one could sample from a branching process model  calibrated with the outage statistics \cite{DobsonPS12} to determine the number of line outages in the next generation and then use the statistics generated in this paper to sample the position on the network  of the line outages. This would give a Monte Carlo way to approximate the extent to which the blackout cascaded beyond the initial damage caused by the quake. While such a method is a rough approximation, it is grounded in the reality of the observed data, and may be a useful approximation in some contexts. For example,  optimized transmission planning  investments accounting for the risk of earthquakes is already highly computationally intensive  \cite{RomeroES15,RomeroPS13}, and a  fast, approximate assessment sampling the effect of cascading would be 
useful\footnote{
Even when it is desirable in principle to model cascading more exactly, there are some computations and contexts in which the computational, modeling and data limitations require a simple stochastic approximation, and it is better to use an approximate model than to omit the effect of cascading entirely.}. 
The problem of estimating the further spread of cascading blackout is particularly important for earthquakes (and other natural disasters or attacks \cite{RomeroPS12}) because earthquakes typically cause much more death, destruction and economic losses than blackouts, but if the response to the earthquake is delayed by a widespread cascading blackout, the losses from the earthquake will be significantly increased.
 \item The statistics of how far cascades typically spread are a starting point for designing 
local and wide area  schemes 
for mitigating cascading. In particular, for design one needs to know typical interaction distances for 
dependent outages (available from probability distributions such as Fig.~\ref{BPACYgenhops}) and the fraction of cascades that are 
confined to the design area of the scheme (available from probability distributions such as Fig.~\ref{BPACYgenmaxhops}). 
\end{itemize}
That is, in addition to validating cascading models and simulations, there are several  promising avenues of engineering risk analysis  that open up given that real cascades can be 
readily tracked spatially on a network.

\section*{Acknowledgement}

We gratefully thank Bonneville Power Administration for 
making publicly available the outage data  that made this paper possible.
The analysis and any conclusions are strictly those of the authors and not of
 Bonneville Power Administration.

 \newpage

 ~
%\vspace{1cm}
 
 \begin{IEEEbiographynophoto}{Ian Dobson} 
 (F'06) received the BA in Mathematics from Cambridge University and the PhD in Electrical Engineering from Cornell University. 
 He previously worked for the British firm EASAMS Ltd and the University of Wisconsin-Madison.
 He  is currently Sandbulte professor of electrical and computer engineering at Iowa State University. 
 \end{IEEEbiographynophoto}

 \begin{IEEEbiographynophoto}
{ Benjamin A. Carreras}
 received the Licenciado en Ciencias degree in physics from the University of Barcelona, Spain and the Ph.D. degree in physics from Valencia University, Spain. He has been a Researcher and a Professor at the University of Madrid, Spain, Glasgow University, U.K., Daresbury Nuclear Physics Laboratory, Warrington, U.K., Junta de Energia Nuclear, Madrid, Spain, and the Institute for Advanced Study, Princeton, NJ. He was a Corporate Fellow at Oak Ridge National Laboratory,  TN. He is now Principal Scientist at BACV Solutions Inc., Oak Ridge, TN. Dr. Carreras is a Fellow of the American Physical Society.
 \end{IEEEbiographynophoto}

 \begin{IEEEbiographynophoto}
{David E. Newman}
 received the B.S. degree in physics and mathematics from the University of Pittsburgh, Pittsburgh, PA and the Ph.D. degree in physics from the University of Wisconsin-Madison. He was a Wigner Fellow and Research Scientist at Oak Ridge National Laboratory,  TN. In 1998, he joined the University of Alaska-Fairbanks, where he is now a Professor of Physics. Dr. Newman is a Fellow of the American Physical Society.
 \end{IEEEbiographynophoto}
 
 \begin{IEEEbiographynophoto}
 {Jos\'e M. Reynolds-Barredo} received the Ingeniero en Telecomunicaciones degree (electrical engineering) from Sevilla University, Spain and the PhD degree in physics from Zaragoza University, Spain. He has been postdoctoral researcher at University of Alaska-Fairbanks, USA. He is currently lecturer at Universidad Carlos III de Madrid, Spain. 
 \end{IEEEbiographynophoto}

\vspace{2cm}
~

\end{document}